\documentstyle{pasj13}
\begin{document}

\SetRunningHead{A.Imada et al.}{photometry of SU UMa}

\title{OAO/MITSuME Photometry of Dwarf Novae: I SU Ursae Majoris} 

\author{Akira \textsc{Imada}$^1$, Hideyuki \textsc{Izumiura}$^1$,
Daisuke \textsc{Kuroda}$^1$, Kenshi \textsc{Yanagisawa}$^1$, \\Toshihiro
\textsc{Omodaka}$^2$, Ryo \textsc{Miyanoshita}$^2$, Nobuyuki
\textsc{Kawai}$^3$, and Daisaku \textsc{Nogami}$^4$}

\affil{$^1$Okayama Astrophysical Observatory, National Astronomical
  Observatory of Japan, Asakuchi,\\ Okayama 719-0232}
\affil{$^2$Faculty of Science, Kagoshima University, 1-21-30 Korimoto,
Kagoshima, Kagoshima 890-0065}
\affil{$^3$Department of Physics, Tokyo Institute of Technology,
  Ookayama 2-12-1, Meguro-ku, Tokyo 152-8551}
\affil{$^4$Kwasan Observatory, Kyoto University, Yamashina-ku, Kyoto 607-8471}
\email{imada@oao.nao.ac.jp}


\maketitle

\begin{abstract}

 We report on simultaneous $g'$, $R_{\rm c}$ and $I_{\rm c}$
 photometry of SU Ursae Majoris during 2011 December - 2012 February
 using OAO/MITSuME. Our photometry revealed that quiescence is divided
 into three types based on the magnitude and color. Quiescent light
 curves showed complicated profiles with various amplitudes and time
 scales. Although no superoutbursts were observed during our run, five
 normal outbursts occurred with intervals of 11 - 21 d. The shapes of
 the normal outbursts were characteristic of the outside-in
 type. During the rising phase of a normal outburst, the light curve
 showed periodic modulations with a period of ${\sim}$ 0.048111(354)
 d, but the origin of this peirod was unclear. We examined daily averaged
 color-color diagram and found that two cycles exist. This
 implies that the thermal limit cycle in SU UMa is complicated. We
 newly discovered that $g'-R_{\rm c}$ becomes red about 3 days prior
 to an outburst. Although the working mechanism on this reddening is
 unclear, we propose two possibilities: one is that the inner portion
 of the accretion disk is filled by matter and obscures the central
 white dwarf, and the other is that the stagnation effect works in the
 outer region of the accretion disk.

\end{abstract}

\section{Introduction}
\begin{table*}[htb]
\caption{Log of observations for MITSuME telescope.}
\begin{center}
\begin{tabular}{cccccccccccc}
\hline\hline
Date & JD(start)$^{*}$ & JD(end)$^{*}$ & Ng$^{\dagger}$ & Nr$^{\dagger}$ & Ni$^{\dagger}$ & Date & JD(start) & JD(end) & Ng & Nr & Ni \\
\hline
2011 Dec. 1 & 897.2545 & 897.3259 & 145 & 142 & 139 & 2012 Jan. 9 & 935.9948 & 936.3805 & 203 & 213 & 199 \\
2011 Dec. 3 & 899.1490 & 899.3622 & 414 & 400 & 414 & 2012 Jan. 11 & 937.9973 & 938.2743 & 549 & 551 & 552 \\
2011 Dec. 4 & 900.0545 & 900.3462 & 477 & 479 & 478 & 2012 Jan. 12 & 939.0511 & 939.3387 & 377 & 382 & 372 \\
2011 Dec. 5 & 901.0121 & 901.3015 & 129 & 136 & 133 & 2012 Jan. 13 & 940.1567 & 940.3839 & 207 & 209 & 206 \\
2011 Dec. 6 & 902.3135 & 902.3448 & 61 & 63 & 61 & 2012 Jan. 14 & 941.0267 & 941.3823 & 167 & 218 & 183\\
2011 Dec. 9 & 905.0096 & 905.1569 & 249 & 252 & 251 & 2012 Jan. 17 & 943.9916 & 944.3802 & 705 & 708 & 703 \\
2011 Dec. 10 & 906.0426 & 906.2374 & 395 & 394 & 395 & 2012 Jan. 21 & 948.1117 & 948.2211 & 221 & 195 & 158 \\
2011 Dec. 11 & 907.1033 & 907.2683 & 121 & 131 & 129 & 2012 Jan. 23 & 950.0767 & 950.3813 & 434 & 455 & 454 \\
2011 Dec. 12 & 907.9916 & 908.3218 & 184 & 240 & 231 & 2012 Jan. 24 & 950.9663 & 951.1166 & 78 & 91 & 80\\
2011 Dec. 13 & 909.0216 & 909.1896 & 209 & 213 & 211 & 2012 Jan. 25 & 952.0032 & 952.3790 & 613 & 610 & 616\\
2011 Dec. 14 & 909.9876 & 910.3193 & 574 & 574 & 562 & 2012 Jan. 26 & 952.9611 & 953.3774 & 504 & 526 & 519\\
2011 Dec. 15 & 911.0066 & 911.0897 & 165 & 167 & 170 & 2012 Jan. 27 & 953.9649 & 954.1563 & 103 & 294 & 274\\
2011 Dec. 16 & 911.9881 & 912.3103 & 468 & 476 & 475 & 2012 Jan. 28 & 954.9791 & 955.2195 & 433& 434& 434\\
2011 Dec. 17 & 912.9852 & 913.3281 & 712 & 713 & 718 & 2012 Jan. 30 & 956.9592 & 957.3757 & 733 & 735 & 734 \\
2011 Dec. 18 & 914.1411 & 914.3215 & 342 & 342 & 342 & 2012 Jan. 31 & 957.9886 & 958.1844 & 414 & 421 & 418 \\
2011 Dec. 19 & 915.0139 & 915.3548 & 678 & 681 & 672 & 2012 Feb. 1 & 958.9539 & 959.2232 & 351 & 355 & 354\\
2011 Dec. 20 & 916.3229 & 916.3198 & 32 & 29 & 28 & 2012 Feb. 2 & 959.9660 & 960.3757 & 596 & 601 & 605\\
2011 Dec. 22 & 918.3166 & 918.3382 & 9 & 9 & 9 & 2012 Feb. 3 & 960.9795 & 961.3458 & 449 & 520 & 465\\
2011 Dec. 23 & 918.9800 & 919.3198 & 342 & 349 & 344 & 2012 Feb. 4 & 962.0366 & 962.3765 & 501 & 527 & 496\\
2011 Dec. 24 & 920.0032 & 920.3043 & 573 & 580 & 580 & 2012 Feb. 7 & 965.0445 & 965.1648 & 187 & 190 & 188\\
2011 Dec. 25 & 921.2649 & 921.3082 & 61 & 63 & 62 & 2012 Feb. 8 & 966.0159 & 966.3652 & 231 & 240 & 232\\
2011 Dec. 26 & 922.0261 & 922.3183 & 483 & 480 & 480 & 2012 Feb. 9 & 967.0471 & 967.3686 & 408 & 428 & 424\\
2011 Dec. 27 & 922.9783 & 923.2906 & 521 & 538 & 529 & 2012 Feb. 10 & 968.0731 & 968.3697 & 279 & 284 & 280\\
2011 Dec. 28 & 924.0758 & 924.1561 & 134 & 134 & 135 & 2012 Feb. 11 & 969.1532 & 969.3714 & 335 & 338 & 338\\
2011 Dec. 29 & 925.0257 & 925.1523 & 180 & 185 & 181 & 2012 Feb. 15 & 973.0054 & 973.3557 & 293 & 306 & 281\\
2011 Dec. 30 & 926.0530 & 926.3813 & 354 & 357 & 354 & 2012 Feb. 16 & 974.0067 & 974.3648 & 246 & 255 & 255\\
2012 Jan. 4 & 930.9585 & 931.1378 & 340 & 339 & 338 & 2012 Feb. 17 & 974.9784 & 975.3640 & 633 & 641 & 636\\
2012 Jan. 5 & 932.0321 & 932.2186 & 365 & 377 & 377 & 2012 Feb. 18 & 976.0219 & 976.3632 & 508 & 510 & 507\\
2012 Jan. 6 & 932.9732 & 933.0944 & 257 & 256 & 255 & 2012 Feb. 19 & 976.9688 & 977.1047 & 280 & 283 & 282\\
2012 Jan. 7 & 933.9873 & 934.3828 & 615 & 616 & 617 & 2012 Feb. 20 & 978.0370 & 978.1697 & 219 & 224 & 220\\
\hline
\multicolumn{12}{l}{$^{*}$JD$-$2455000 $^{\dagger}$Number of available
 data in $g'$, $R_{\rm c}$, and $I_{\rm c}$ bands.} \\
\end{tabular}
\end{center}
\label{obslog}
\end{table*}
Cataclysmic variables are semi-detached interacting binaries that
consist of a white-dwarf primary and a late-type secondary. The
secondary star fills its Roche lobe, transferring mass into the primary
Roche lobe via the inner Lagrangian point (L1), by which the accretion
disk is formed around the primary (for a review, see
\cite{war95book}; \cite{hel01book}). Dwarf novae constitute a subclass
of cataclysmic variables, further classified into three subclasses
according to their activities (for a review, see \cite{osa96review};
\cite{osa05review}). SU UMa-type dwarf novae are one
subclass of dwarf novae. They show two types of outbursts: normal
outburst which lasts for a few days and superoutburst which lasts for
more than 10 days \citep{kat04vsnet}. During the superoutburst,
tooth-like modulations termed superhumps are visible. The mean period
of the superhump are a few percent longer than that of the orbital
period, which are believed to be caused by phase-dependent tidal
dissipation of the eccentrically-deformed precessing accretion disk
(\cite{whi88tidal}; \cite{hir90SHexcess}).

It is widely accepted that the mean superhump period varies as the
superoutburst proceeds. This indicates that the structure of the accretion
disk varies, or the eccentric mode propagates across the disk, or
both \citep{uem05tvcrv}. Recently, statistical studies of superhump
period changes were performed by T. Kato and his colleagues, in which they
calculated superhump period changes for as large as 500
superoutbursts (\cite{pdot}; \cite{pdot2}; \cite{pdot3};
\cite{pdot4}). Such an unprecedented statistical studies have revealed
the basic picture of the superhump period change. The textbook of
$O-C$ diagram consists of three stages: stage $A$ when the mean
superhump period is constant, stage $B$ when the mean superhump period
increases as the superoutburst proceeds, and stage $C$ when the mean
superhump period keeps constant but the period slightly shorter than
that of stage $A$. The vast majority of SU UMa-type dwarf novae
follows the basic picture of $O-C$ diagram. However, there exist
some exceptions. For instance, some objects do not show stage
$A$. Another exception is that the mean superhump period
decreases during stage $B$. In order to understand the observed
diversity of the superhump period changes, time-resolved photometry
not only during superoutburst but also during quiescence are
imperative.

Another noticeable topic is that unprecedentedly precise light
curves with one minute cadence have been provided by $Kepler$ satellite
(\cite{kepler1}; \cite{kepler2}). $Kepler$ observations include some
dwarf novae (\cite{2010ApJ...717L.113S}; \cite{woo11v344lyr};
\cite{2012MNRAS.422.1219B}; \cite{osa13v1504cyg}). Especially, V344
Lyr and V1504 Cyg are well studied. Thanks to $Kepler$ observations,
researchers have been able to analyze
sophisticated data ever obtained. For example, \citet{2010ApJ...725.1393C}
and \citet{2012ApJ...747..117C} examined various parameters of each
outburst and intervals of quiescence, and conclude that the thermal-tidal
instability model faces a difficulty in explaining quiescent intervals
between normal outbursts during a supercycle. On the other hand,
\citet{osa13v1504cyg} investigated $Kepler$ data of V1504 Cyg and
concluded that the overall light curve is well reproduced by the
thermal-tidal instability model. From an observational viewpoint,
\citet{pdot3} detected superhumps during a normal outburst just prior
to the superoutburst of V1504 Cyg. \citet{pdot3} also studied
superhump period changes of these objects. In V344 Lyr, \citet{pdot3}
reported that a stage B$-$C transition may be associated with the
secondary component of the superhumps that emerges from the middle of
the plateau phase.

\begin{table}
\caption{Log of observations for Kagoshima 1m telescope.}
\begin{center}
\begin{tabular}{ccccc}
\hline\hline
Date & JD(start)$^{*}$ & JD(end)$^{*}$ & exp$^{\dagger}$ & N$^{\ddagger}$ \\ 
\hline
2012 Mar. 13 & 1000.1499 & 1000.2802 & 45 & 194 \\
2012 Mar. 14 & 1001.0624 & 1001.2710 & 30 & 300 \\
\hline
\multicolumn{4}{l}{$^{*}$JD$-$2455000.} \\
\multicolumn{4}{l}{$^\dagger$ Exposure time in units of seconds.} \\
\multicolumn{4}{l}{$^\ddagger$ Number of exposure.} \\
\end{tabular}
\end{center}
%\label{sites}
\end{table}

$Kepler$ data of V344 Lyr and V1504 Cyg provide further scientific
topic: study of negative superhumps. Negative superhumps
are considered to be a consequence of retrograde precession of the
tilted accretion disk (\cite{pat95v1159ori}; \cite{2009ApJ...705..603M};
\cite{2012ApJ...745L..25M}). However, physical mechanisms for
generating negative superhumps are still in debate
\citep{2012ApJ...753L..27M}. In general, negative superhumps are known to be
observed occasionally in ER UMa-type and novalike stars (\cite{pat95v1159ori};
\cite{2012NewA...17..433R}; \cite{oht12eruma}). However, recent
analyses reveal that negative superhumps are detected not only in
these systems, but also in WZ Sge-type stars and typical SU UMa-type stars
(\cite{2010ApJ...717L.113S}; \cite{pdot2};
\cite{pdot3}). From the theoretical side, \citet{2012ApJ...753L..27M}
performed SPH simulations and succeeded in reproducing accretion disks
in SU UMa-type dwarf novae that tilt, warp, and precess in the
retrograde direction. She concludes that negative superhumps and/or
their signals should be ubiquitous below the period gap. Using the
frequencies of negative superhumps, \citet{osa13v1504cyg} estimated
the disk radius of the accretion disk of V1504 Cyg and confirmed that
it varies as predicted by the thermal-tidal instability model.

In response with the above mentioned results, we set up multicolor
photometry using OAO/MITSuME \citep{3me}. Our goals are, to constrain
the physical mechanism that causes the diversity of $O-C$ diagram of
SU UMa-type dwarf novae, to understand quiescent light curves and
color indices, and to understand the nature of negative superhumps. The
present photometric campaign provides an unprecedentedly long-term
light curve of SU UMa itself. Although no superoutbursts were observed
during our run, we obtained extensive light curves and color
indices. Part of our results has been already published as a letter
\citep{ima12suumaletter}.

\section{Observations}

Simultaneous $g'$, $R_{\rm c}$, and $I_{\rm c}$ photometry were
performed from 2011 December 1 to 2012 February 20 using the
50cm-MITSuME telescope located on Okayama Astrophysical Observatory
\citep{3me}. A journal of observations is given in table
\ref{obslog}. The total datapoints of our run exceed 60000, which is
the largest photometric campaign ever performed for SU
UMa. On 2012 January 21 and 23, the shutter of CCD in $g'$ band did
not work very well, so that we excluded these data for period and color
analyses.

After excluding bad data, we used 20191, 21459, and 20975 datapoints for
$g'$, $R_{\rm c}$, and $I_{\rm c}$ bands, respectively. The obtained
images were analyzed with aperture photometry using
IRAF/daophot\footnote{IRAF (Image Reduction and Analysis Facility) is
  distributed by the National Optical Astronomy Observatories, which
  is operated by the Association of Universities for Research in
  Astronomy, Inc., under cooperative agreement with National Science
  Foundation.}. We derived magnitudes and colors of SU UMa
with differential photometry using Tycho-2 4126-00036-1 (RA:
08:12:45.104, Dec: +62:26:17.57), whose constancy was checked by nearby
stars in the same images. We adopted $g'$=10.72, $R_{\rm c}$=10.41, and
$I_{\rm c}$=9.98 as the magnitudes of the comparison star. These values
may contain some errors, but such uncertainty will not influence on the
main results, because we focus on the variations of light curves and
avoid estimation of temperatures based on color indices.

We also performed $J$ band photometry using the 1m-telescope of Kagoshima
university. Table 2 shows a log of $J$ band photometry. Exposure times
were 30 sec for 2012 March 13 and 45 sec for 2012 March 14,
respectively. To obtain better S/N, we summed two consecutive frames, so
that substantial exposure times were 60 sec and 90 sec,
respectively. Because of the narrow view of the telescope
(5'$\times$5'), we cannot use the same
comparison star as that used in the optical. Instead we adopted 2MASS
J081237.88+623842.3 ($J$=13.039) as the comparison star, whose constancy
was checked by 2MASS J081232.76+623743.5 ($J$=14.193) and 2MASS
J081247.05+623700.0 ($J$=13.562). Unfortunately, simultaneous
optical-near infrared photometry were unable to be performed because
another observational schedule ran on MITSuME telescope .

Heliocentric correction was made before the following analyses.

\section{Results}

\begin{figure*}[htb]
\begin{center}
\FigureFile(160mm,100mm){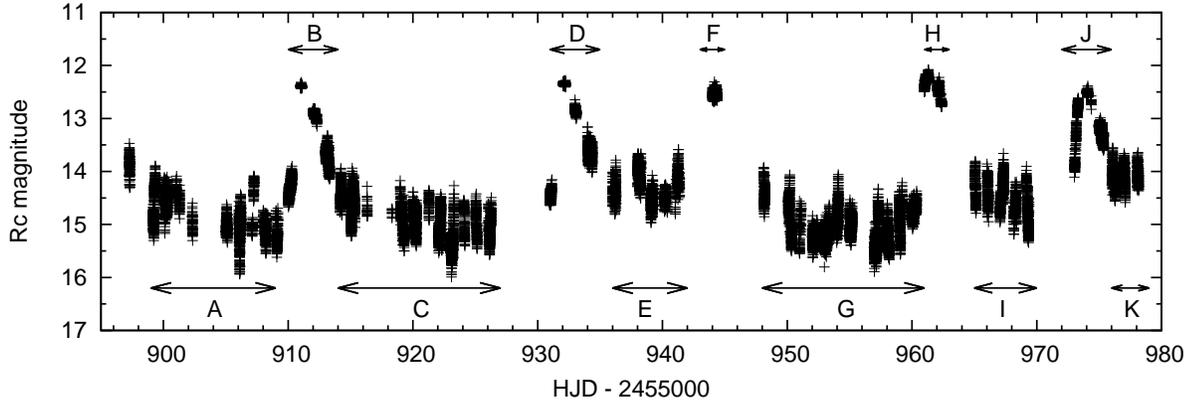}
\end{center}
\caption{Overall light curve of our observations. The abscissa and
 ordinate denote HJD$-$2455000 and $R_{\rm c}$ magnitude, respectively. During
 our observations, SU UMa exhibited 5 normal outbursts. Quiescence can
be divided into two types according to averaged magnitudes (stage A,
C, G and stage E, I, K). On HJD 2455907, a flare-like event was
observed, during which the light curve varied at a rate of $-$6.1
mag/d.}
\label{lc}
\end{figure*}

\begin{figure*}[htb]
\begin{center}
\FigureFile(160mm,80mm){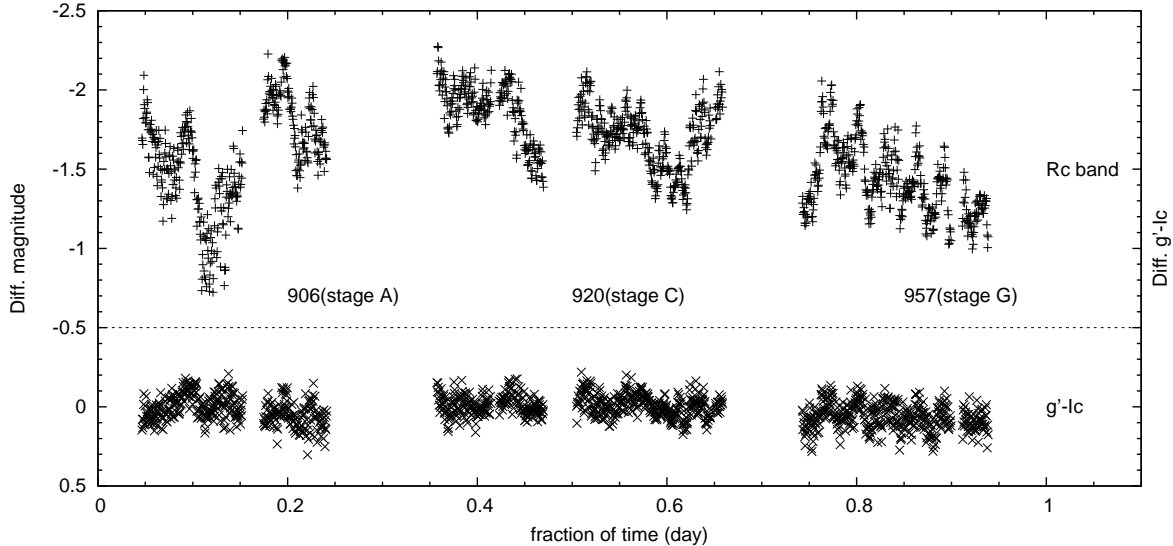}
\end{center}
\caption{Representative $R_{\rm c}$ light curves (top) and
 $g'-I_{\rm c}$ variations (bottom) during faint quiescence. Numbers
  in the figure denote the days since HJD 2455000. Each
 light curve shows complicated profiles with various time scales. No
 coherent variations are found in the light curves. The color indices in
 $g'-Ic$ are associated with $R_{\rm c}$ magnitudes: when the $R_{\rm c}$
 magnitudes get brighter, the color indices get bluer.}
\label{a_rep}
\end{figure*}

\begin{figure*}[htb]
\begin{center}
\FigureFile(160mm,80mm){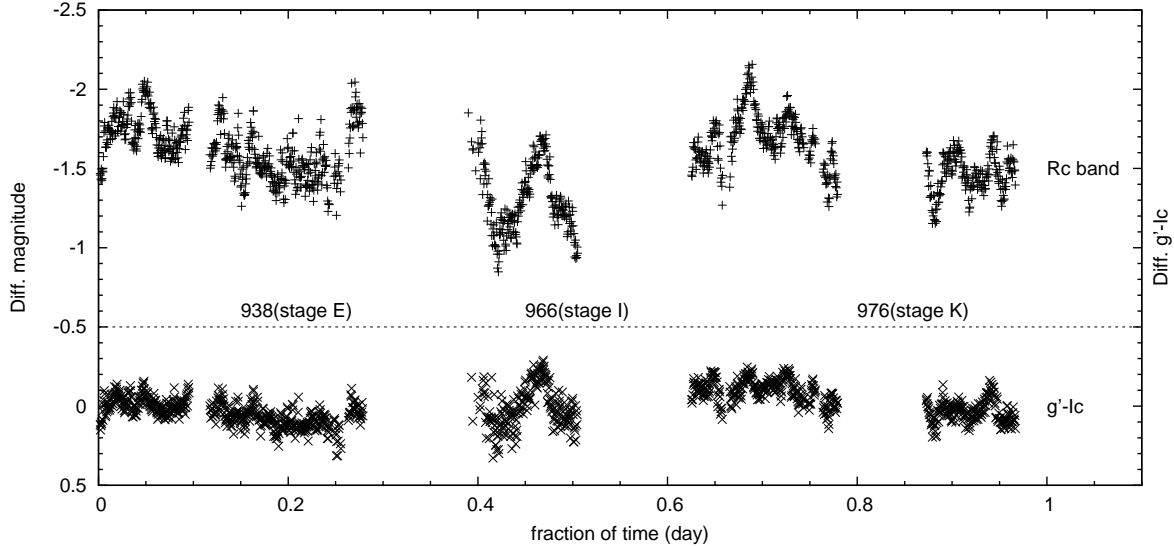}
\end{center}
\caption{Same as figure 2 but during bright quiescence. The relation
  between $R_{\rm c}$ and $g'-I_{\rm c}$ is the same as that during
  faint quiescence.}
\label{bqui_rep}
\end{figure*}

\begin{table}[htb]
\caption{Candidate periods during quiescence.}
\begin{center}
\begin{tabular}{cc}
\hline\hline
Period(err) & stage \\
\hline
0.074108(23) & C \\
0.076838(103) &C \\
0.072217(76) & E \\
0.075021(54) & E \\
0.077971(59) & E \\
0.077650(35) & G \\
0.072635(52) & I \\
0.077876(59) & I \\
0.072401(96) & K \\
0.075127(86) & K \\
0.078218(133) & K \\
\hline

\hline
%\multicolumn{4}{l}{$^\dagger$ Exposure time in units of seconds.} \\
%\multicolumn{4}{l}{$^\ddagger$ Number of exposure.} \\
\end{tabular}
\end{center}
\label{cand}
\end{table}

\begin{figure}[htb]
\begin{center}
\FigureFile(80mm,){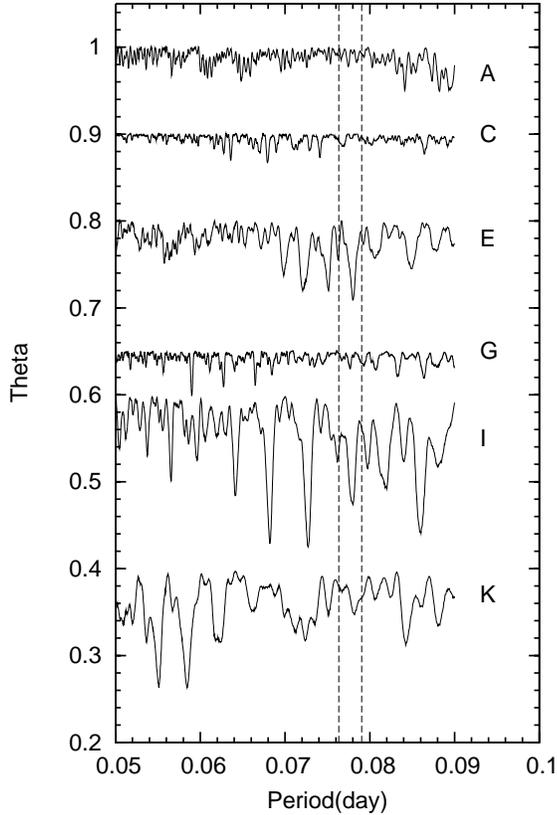}
\end{center}
\caption{Theta diagrams during quiescence. The letters in this figure
 denote the stages displayed in figure 1. The vertical dashed lines
 correspond to $P_{\rm orb}$ = 0.07635 d and $P_{\rm sh}$ = 0.07875 d,
 respectively.}
\label{qui_pdm}
\end{figure}

\begin{figure}[htb]
\begin{center}
\FigureFile(80mm,80mm){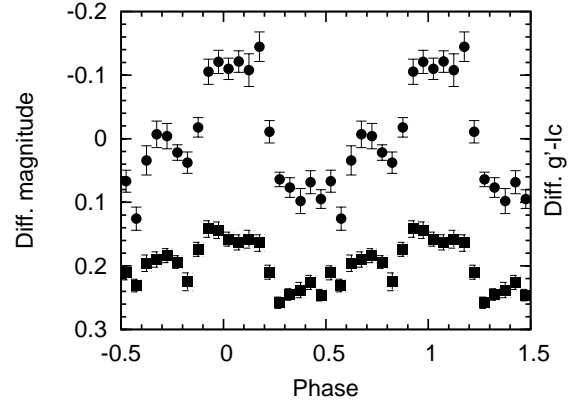}
\end{center}
\caption{An example of phase-averaged $R_{\rm c}$ band light curve
  (top, filled circles) and $g'-I_{\rm c}$ (bottom, filled squares)
  folded with 0.072635 d. Although the light curve shows hump-like
  modulations, no robust evidence for positive and negative superhumps
is provided.}
\label{i_r_g-i}
\end{figure}

\begin{figure}[htb]
\begin{center}
\FigureFile(80mm,80mm){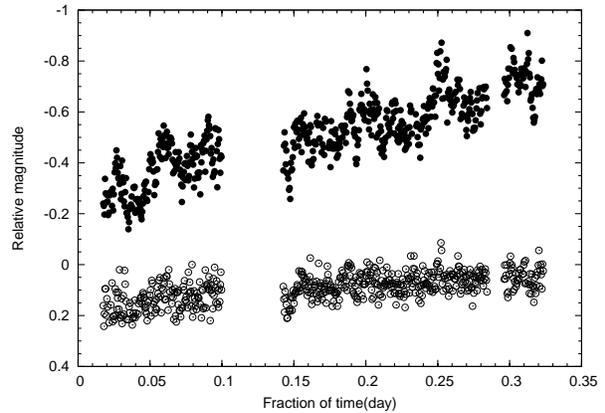}
\end{center}
\caption{$R_{\rm c}$ light curve (top: filled circles) and $g'-I_{\rm c}$ variation (bottom: open circles) on
  2011 December 14, at the very onset of a normal outburst. Hump-like
  modulations with an amplitude of ${\sim}$ 0.3 mag are visible in the
  light curve.} 
\label{1214}
\end{figure}

\begin{figure}[htb]
\begin{center}
\FigureFile(80mm,80mm){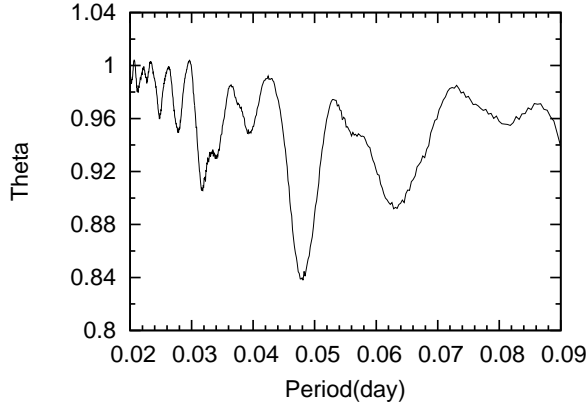}
\end{center}
\caption{PDM analysis for the light curve on 2011 December 14 after
  removing the trend. The strongest signal corresponds to
  $P$=0.048111(354) d. This periodicity can be checked with simple eye
  estimation of the light curve.}
\label{1214pdm}
\end{figure}

\begin{figure}[htb]
\begin{center}
\FigureFile(80mm,80mm){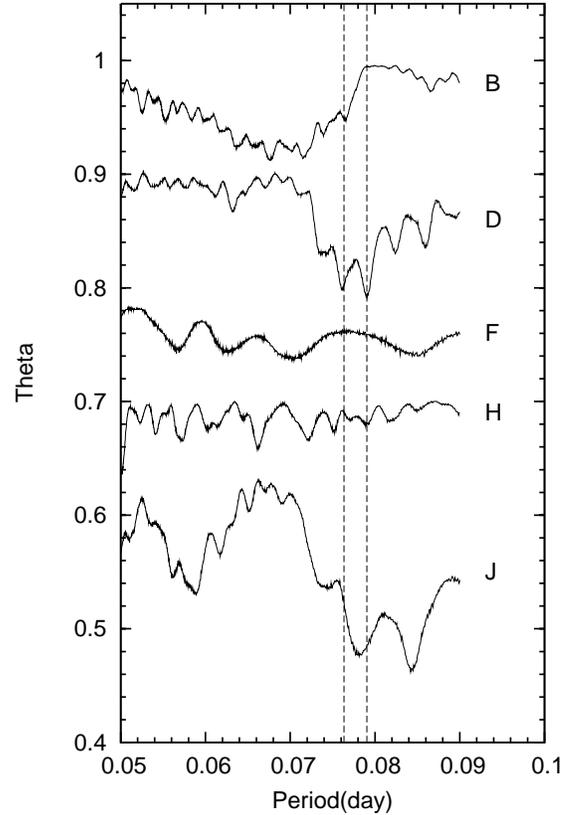}
\end{center}
\caption{PDM analyses during normal outbursts. The letters and
  dashed lines are the same as those of figure \ref{qui_pdm}. The
  theta diagram of stage D coincide with figure 2 of
  \citet{ima12suumaletter}.}
\label{out_pdm}
\end{figure}

\begin{figure}[htb]
\begin{center}
\FigureFile(80mm,80mm){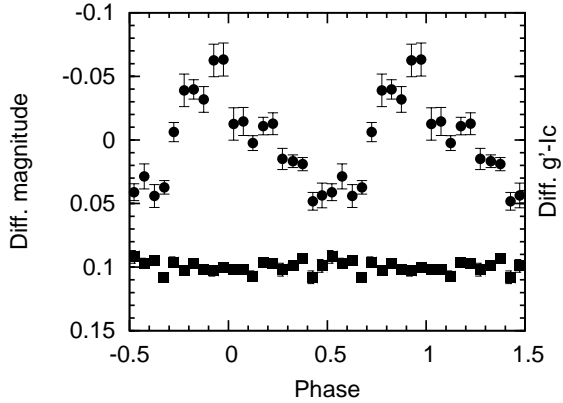}
\end{center}
\caption{Phase-averaged $R_{\rm c}$ band light curve (top, filled circles) and
$g'-I_{\rm c}$ (bottom, filled squares) folded with 0.078365 d. A rapid rise and
slow decline, reminiscent of superhumps, are visible in the light curve. On the
other hand, no significant variations are visible in $g'-I_{\rm c}$.}
\label{j_r_g-i}
\end{figure}

\begin{figure}[htb]
\begin{center}
\FigureFile(80mm,80mm){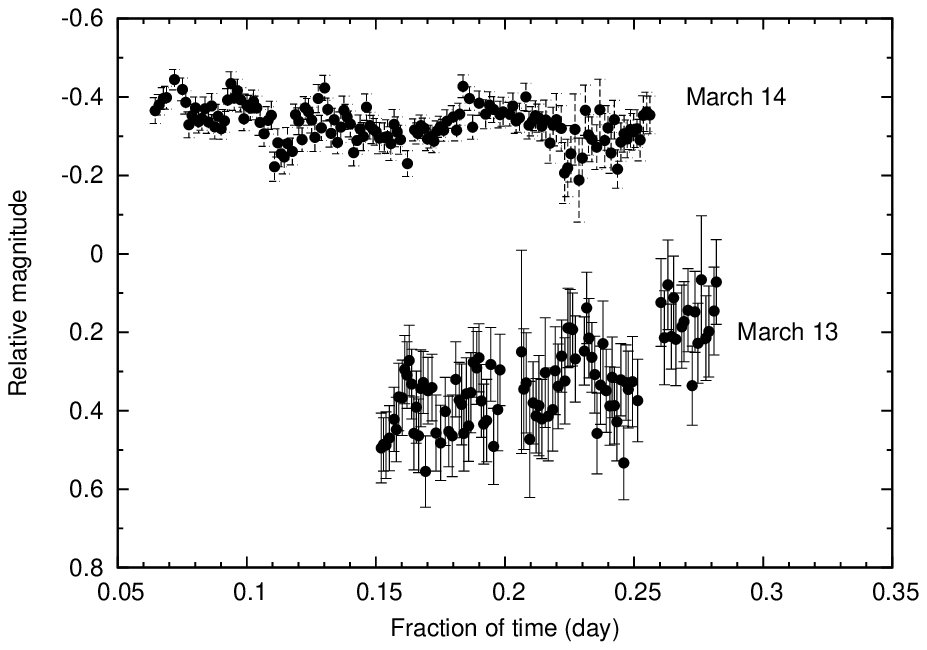}
\end{center}
\caption{$J$ band light curves of SU UMa obtained with KU 1m. The x and
 y axes denote fractional time (unit in day) and relative magnitude,
 respectively. Although the datapoints contain large error bars,
 hump-like modulations are detectable in both light curves. Each
 datapoint is arbitrarily shifted for display purpose.}
\label{ku_j}
\end{figure}

\subsection{Overall light curve}

Figure \ref{lc} shows the obtained light curve of $R_{\rm c}$
band. During our observations, SU UMa exhibited five normal outbursts at
intervals of 11-21 days. According to the AAVSO light curve
generator, a normal outburst occurred on 2011 November
30\footnote{www.aavso.org/lcg}. Based on this observation, we safely
regard slight brightness on 2011 December 1 as the declining stage of
the normal outburst. It should be
noted that the quiescent magnitudes in stage E, I and K are obviously
brighter by ${\sim}$ 0.5 mag than those in other stages. We call stage 
E, I, and K ``bright quiescence'', and stage A, C, and G ``faint quiescence'' 
throughout this paper. In addition, we
briefly note that a flare-like event was observed on 2011 December
11, when $R_{\rm c}$ magnitude varied from 15.5 to 14.5 at a
rate of $-$6.1(1) mag/d. Similar phenomenon was reported by
\citet{ech96suumaminioutburst} in which the magnitude changed
drastically within a time scale of ${\sim}$ 0.2 d (see figure 1a of
\citet{ech96suumaminioutburst}).

\subsection{Quiescence}

Figure \ref{a_rep} and \ref{bqui_rep} displays representative light curves and 
$g'-I_{\rm c}$ variations during faint and bright quiescences, respectively. 
As can be seen in these figures, the light curves show incoherent profiles 
with various time scales and amplitudes. It also should be noted that the
maxima of the magnitude coincide with the blue peaks of $g'-I_{\rm
  c}$. This is in contrast with that observed during superhumps,
during which the bluest peak in $g'-I_{\rm c}$ is prior to $R_{\rm c}$
by a phase of ${\sim}$ 0.2 (\cite{mat09v455and}; \cite{ima12suumaletter}).

In order to search for periodic signals in the light curves, we performed 
the phase dispersion minimization method (PDM, \cite{ste78pdm}) for detrended 
light curves. Figure \ref{qui_pdm} shows results of PDM
analyses. As can be seen in this figure, these theta diagrams imply
many periodicities. In this section, we focus on periods around the
orbital period of the system ($P_{\rm orb}$=0.07635 d,
\cite{tho86suuma}). Table \ref{cand} lists candidate periods around the
orbital period. In order to clarify the origin of these periodicities,
we folded light curves and $g'-I_{\rm c}$ variations with the periods
listed in table \ref{cand}. An example of the results is shown in
figure \ref{i_r_g-i}, in which hump-like modulations are
visible. However, we could not find robust evidence for negative
superhumps, whose profiles show a slow rise and rapid decline in their
light curves \citep{osa13v1504cyg}. Regarding positive superhumps, a
hint of them was shown in some folded light curves. But no
phase discordances between light curves and colors were observed.

\subsection{Outburst}

During our run, SU UMa experienced five normal outbursts, of which three
(stage B, D, and J) were well observed. The overall profiles
of these outbursts resemble those of type-A (outside-in) outburst
\citep{sma84DI}. The declining rates of these normal outbursts were
estimated to be ${\sim}$ 1 mag/d, typical for those observed in SU
UMa-type dwarf novae \citep{ole04ttboo}.

We performed PDM analyses during the normal outbursts. Figure
\ref{out_pdm} shows the resultant theta diagrams applied to the fading
stages of each normal outburst. During stage D, we reported on
detection of superhumps in the previous letter
\citep{ima12suumaletter}. As for stage J, a hint of superhumps was
observed with the period of $P$=0.078365(295) d. We folded the light
curve and $g'-I_{\rm c}$ color with this period. Figure \ref{j_r_g-i}
exhibits the resultant data. Although the phase averaged light curve
is reminiscent of superhumps, no significant color variations are found.

Figure \ref{1214} illustrates the enlarged light curve and color on
2011 December 14, corresponding to the onset of the outburst (stage
B). One can notice cyclic variations with amplitudes of
${\sim}$ 0.3 mag. During the rising stage, the magnitude brightened at
a rate of $-$1.4(1) mag/d. This value is unusually smaller compared
with those observed in other SU UMa-type dwarf novae ($-$6 mag/d,
\citet{war95book}). Judging from the obtained value, the rising rate may
have changed after our observation. After removing the rising trend,
we performed a period analysis for the residual light curve. The
resultant theta diagram is displayed in figure \ref{1214pdm}, from
which we derived $P$=0.048111(354) d as a candidate period. This
periodicity can be checked by simple eye-estimation of the light
curve. We also tried to examine the rising phase of stage J, but noisy
data prevented us from a period analysis.

\subsection{J band photometry}

Figure \ref{ku_j} gives whole $J$ band light curves of SU UMa. Although
the data were acquired under low S/N, hump-like modulations with
an amplitude of ${\sim}$ 0.2 mag were visible. The mean magnitudes
were estimated as $J$=13.36 mag for 2012 March 13 and $J$=12.11
mag for 2012 March 14, respectively. A hint of a rising trend was
observed on 2012 March 13, when the magnitude varied at a
rate of $-$1.9(1) mag/d. According to the AAVSO light curve generator, the
visual magnitudes yield $V$=14.0 on 2012 March 13, $V$=12.3 on March
14, and $V$=14.6 on 2012 March 18, respectively. These results indicate
that our $J$ band photometry were performed at the onset of the normal
outburst.

\section{Discussion}

\subsection{color-color diagram}

\begin{figure*}[htb]
\begin{center}
\FigureFile(160mm,160mm){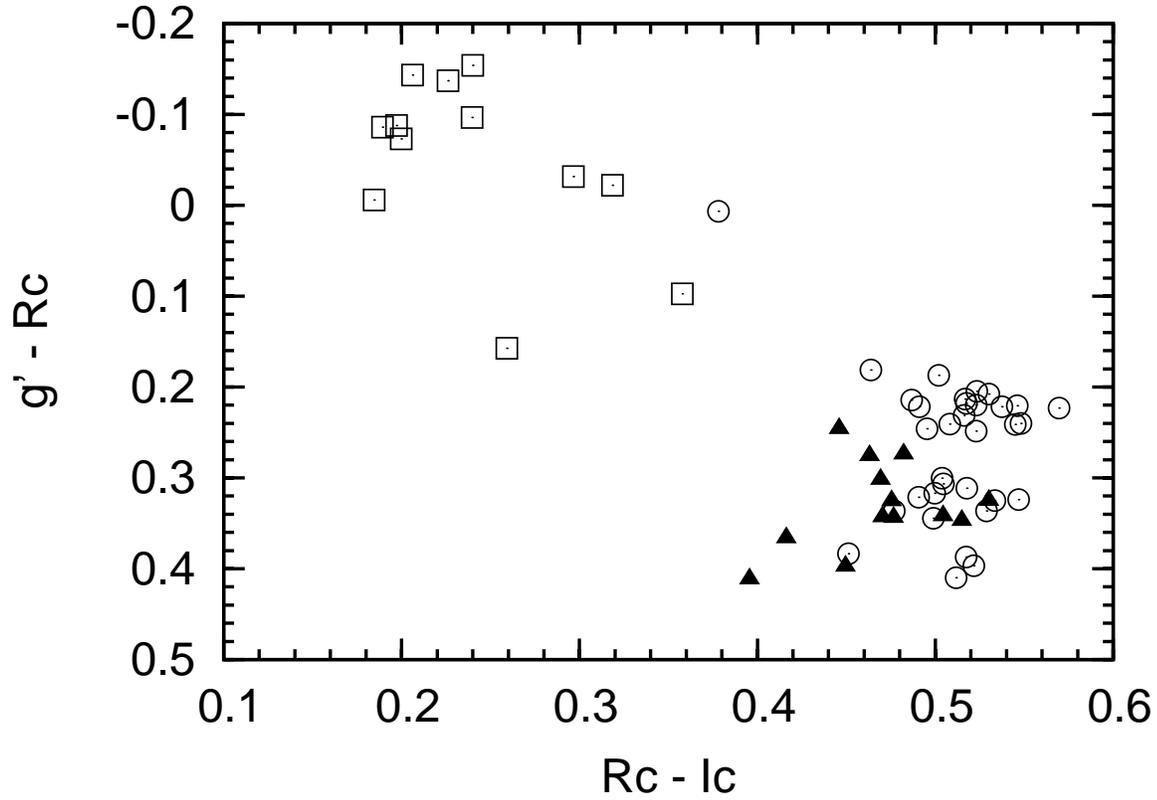}
\end{center}
\caption{Daily averaged color-color diagram. Open
 circles, filled triangles and open squares denote datapoints during
 faint quiescence, bright quiescence, and outburst, respectively. Note
 that there exists the ``gap'' around $g'-R_{\rm c}$ $\sim$ 0.28. Note that all
 of datapoints during bright quiescence are located below the gap.}
\label{col_a}
\end{figure*}

Figure \ref{col_a} displays the daily averaged color-color diagram of SU
UMa. As can be seen in this figure, bright quiescence tends to show redder
$g'-Rc$ and bluer $Rc-Ic$ compared with those of faint quiescence. In
addition, one can notice the ``gap'' at a threshold of $g'-Rc$
$\sim$ 0.28, which indicates that the magnitude in $g'$ band becomes
more faint than that in $Rc$ and $Ic$ bands. It also should be noted that
all datapoints of $g'-Rc$ during bright quiescence are located in the
red side of the gap.

\begin{figure*}[htb]
\begin{center}
\FigureFile(160mm,160mm){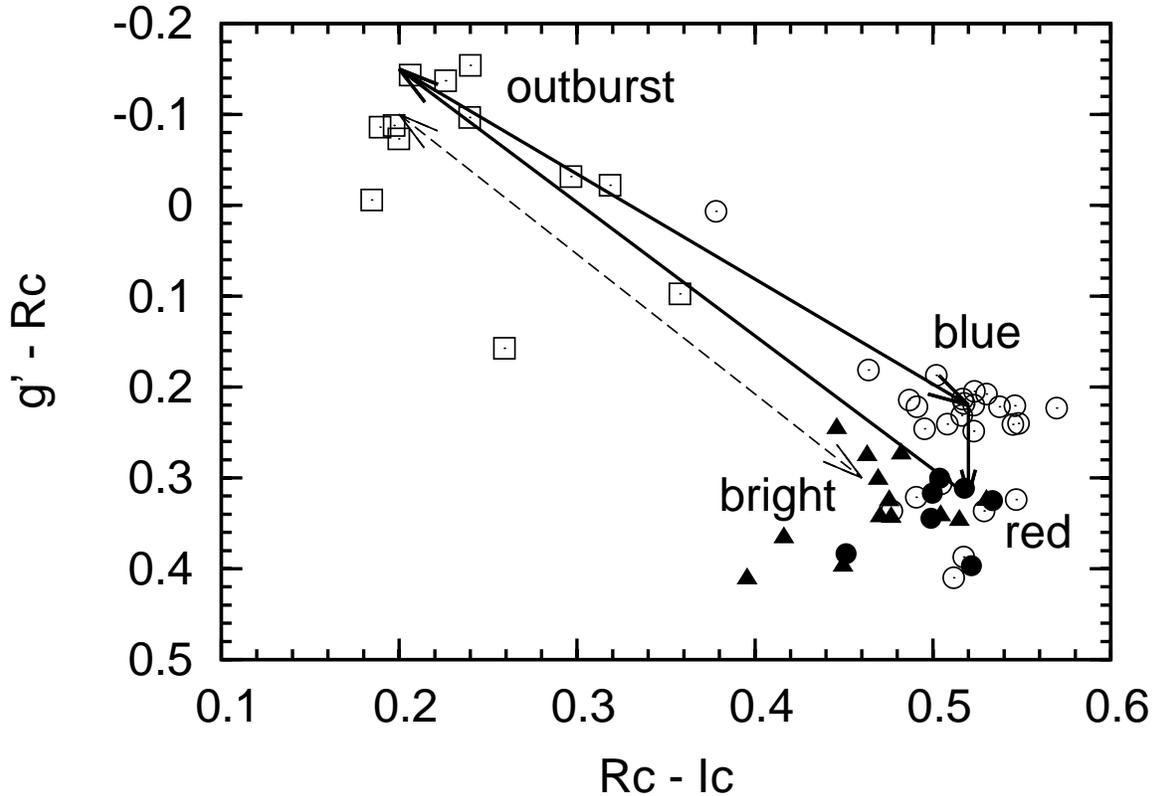}
\end{center}
\caption{Same diagram as figure \ref{col_a} but we denote datapoints
  3 day prior to outburst as filled circles. Note that all of the
  filled circles are plotted below the gap, which means that
  $g'-R_{\rm c}$ becomes redder prior to outburst. We draw possible
  cycles of color variations with solid and dashed arrows, respectively.}
\label{col_b}
\end{figure*}

In order to further understand the color-color diagram, we investigated
each datapoint of faint quiescence. We found that, (1) after the end of
an outburst, color indices of $g'-Rc$ tend to be located in the blue
side of the gap although there exist exceptions, and (2) from a
few days prior to outburst, color indices of $g'-Rc$ are located in the
red side of the gap without exceptions. Taking these facts into
account, we can draw the evolutions of color indices, which are given in
figure \ref{col_b}. Based on this diagram, we can find
two cycles: outburst $\rightarrow$ faint quiescence with bluer
$g'-R_{\rm c}$ $\rightarrow$ faint quiescence with redder $g'-R_{\rm c}$
$\rightarrow$ outburst (marked with solid arrows) and outburst
$\rightarrow$ bright quiescence $\rightarrow$ outburst (marked with a
dashed arrow). The absence of data in the gap implies that
a typical time scale of the transition from blue to red side is as
short as 1 day. The coexistence of the cycles suggest that the thermal
equilibrium curve of the disk instability model is more complicated
such as figure 2 of \citet{min85DNDI}, compared with that of the
simple $S$-shaped curve.

Recently, \citet{pri07sscyg} performed extensive $V$ and $I_{\rm c}$
photometry of the prototypical dwarf nova SS Cyg and reported that a
hint of a variation in the $V-I_{\rm c}$ color was observed 5 days prior to
outburst. Although we used $g'$, $R_{\rm c}$ and $I_{\rm c}$ bands, we
confirmed a color variation prior to outburst, as suggested by
\citet{pri07sscyg}. In the present study, we have found that, at least SU
UMa itself, the values of color indices are a powerful indicator for
expecting an impending outburst.

Why is $g'$ band faint before an outburst? One possibility is that
inner portion of the accretion disk is filled by matter, which obscures
the flux from the white dwarf. If this is the case, then line profiles
in Balmer series such as H$\beta$ and H$\gamma$ should change before an
outburst; wing velocities will increase because of the existence of the
inner matter. Another possibility is that the stagnation effect,
originally propounded by \citet{min88uvdelay}, works in the accretion
disk. \citet{min88uvdelay} predicts that the outer cool region makes a
halt in a warm state with $T_{\rm eff}$ ${\sim}$ 6000 $K$ at the onset
of type-A outbursts. If the stagnation stage occurs prior to outburst,
this can bring $g'-R_{\rm c}$ to a redder side of color-color
diagram. In order to clarify the nature of the faintness in $g'$ band,
spectroscopic observations prior to outburst should be performed.

\subsection{Quiescence}

One of the most significant findings in our observations is that we
confirm the presence of bright quiescence in SU UMa itself. It is well
known that, after the end of superoutburst, SU UMa-type dwarf novae
keeps brighter magnitudes compared with those of the majority of
quiescence
(\cite{pat98egcnc}; \cite{ste07wxcet}). This trend can be also seen in
the $Kepler$ data of V344 Lyr and V1504 Cyg, although there exist some
exceptions (\cite{2010ApJ...725.1393C};
\cite{osa13v1504cyg}). \citet{2010ApJ...725.1393C} noted that this is
associated with cooling of the white dwarf heated by a superoutburst
\citep{sio95CVWDheating}, although the slow decay in V344 Lyr for
about 50 days after a
superoutburst mentioned in Cannizzo et al. (2010) might be due to a
purely instrumental effect accompanied by repositioning of $Kepler$
spacecraft (see \S 2.1 in
\citet{2012ApJ...747..117C}). \citet{2011AJ....141...84B} suggest that
the white dwarf of GW Lib was heated by the 2007 superoutburst and has
not returned to the quiescent state. Therefore one can expect that
appearance of bright quiescence is restricted in the vicinity of the
termination of the superoutburst.

The present observations, however, show that appearance of bright
quiescence is independent of superoutburst. In addition, the durations
of bright quiescence tend to be shorter compared to those of faint
quiescence. Although the reason for entering bright quiescence should
await further observations, it may attribute to the accretion disk
itself, rather than the heated white dwarf.

Regarding light curves, they show complicated profiles with various time
scales and magnitudes. As mentioned the previous section, no
significant signals corresponding to the orbital period were detected
in the PDM analyses. This implies that the main light source
may be originated from a complicated structure of the accretion disk,
rather than the hot spot. Color variations and light curves correlate
well each other: the maxima of the light curves coincide with the
blue peaks of $g'-I_{\rm c}$. This differs from those observed in
superhumps, in which the blue peaks of $g'-I_{\rm c}$ is prior
to the bright maxima of $R_{\rm c}$ band by phase ${\sim}$ 0.2
(\cite{mat09v455and}; \cite{ima12suumaletter}). \citet{mat09v455and}
noted that such behavior is associated with the heating and expansion
in the accretion disk. From this standpoint, the present result implies
that these processes are marginal during quiescence.

\subsection{outburst}

As noted in the previous section, SU UMa exhibited five normal
outbursts during our observations. The bright maxima occurred on 14,
35, 46, 63, and 77 d from the epoch of our observations, by which we
roughly estimated recurrence times as 21, 11, 17, and 14 d. According
to the original thermal-tidal instability, the recurrence time of normal
outburst increases with the supercycle phase
\citep{osa89suuma}. However, the present observations show that no
regularity exists in the recurrence times of the normal
outbursts. Similar trend can be also seen in the $Kepler$ light curves
of V344 Lyr and V1504 Cyg, in which the duration of quiescence ranges
from 3 d to 20 d. So far, our obtained data cannot lead to a conclusion
why SU UMa shows various durations of quiescence.

As displayed in figure \ref{1214}, we succeeded in observing the very
onset of the outburst, in which the light curve showed hump-like
modulations with a period of 0.048111(354) d. If these modulations are
linked to the Keplerian motion of the accretion disk, then the
light source is originated from ${\sim}$ 0.7$a$, where $a$ denotes the
binary separation. This value is far beyond the tidal truncation radius of
the accretion disk of SU UMa. It is unlikely to be
the case either that these are associated with the hot spot, since no
sign of the orbital period is detected. At present, the origin of the
modulations and periodicity is unclear. In order to clarify the
physical mechanism of these modulations, further photometric data
before the onset of an outburst should be collected.

\subsection{negative superhumps}

One of the purposes of our observations is to study negative
superhumps of SU UMa-type dwarf novae. As noted in the previous
section, our PDM analyses imply the signals shorter than the orbital
period of the system. \citet{osa13v1504cyg} noted that negative
superhumps show coherent light curves with a slow rise and rapid
decline, like figure 8 of their paper. \citet{woo11v344lyr} reported
that appearance and disappearance of negative superhumps may be
triggered by an outburst, based on period analyses on V344 Lyr. In
conjunction with these results, it may be premature to conclude that
we detect negative superhumps, even though the theta diagrams imply
the periodicities shorter than the orbital period.

Although we are less confident on detection of negative superhumps,
our result does not mean that SU UMa shows any negative superhumps at
all. For example, V344 Lyr and V1504 Cyg
exhibited negative superhumps for only 150 d and 260 d against more
than 600 d data (\citet{woo11v344lyr}; \citet{osa13v1504cyg}). Taking
these facts into consideration, our observations are too short to
conclude whether SU UMa shows negative superhumps.

As noted in \citet{ima12suumaletter}, SU UMa experienced unusual states
in the past (\citet{ros00suuma}; \citet{kat04rxandsuuma}). According
to \citet{ros00suuma}, SU UMa hardly showed outbursts for about 3
years, despite the fact that a mean supercycle of SU UMa being ${\sim}$ 1
year. In addition, SU UMa occasionally entered long quiescence with
durations of longer than 50 d \citep{ros00suuma}. Recently, many
authors point out that appearance of negative superhumps suppress an
outburst (\citet{oht12eruma}; \citet{osa13v1504cyg}). Based on these
ideas, we speculate that such a long quiescence might be associated
with negative superhumps. This should be investigated by further
quiescent photometry not only for SU UMa itself, but also for other SU
UMa-type dwarf novae.

\section{Summary}

In this paper, we summarize our results as follows:

1. Quiescence is divided into two types according to the magnitude:
faint quiescence and bright quiescence. The latter is further divided
into two types based on $g'-R_{\rm c}$ color. During bright
quiescence, the averaged magnitude is $\sim$ 0.5 mag brighter than
that of faint quiescence.

2. The obtained light curves varied with the amplitude as
large as $\sim$ 1 mag with irregular and incoherent profiles. Although
our period analyses imply the periodicities shorter than the orbital
period, it is premature to conclude that we detect negative superhumps.

3. During our photometric campaign, SU UMa experienced five normal
outbursts, of which three were successfully observed. These outbursts are
characteristic of the outside-in type. At the rising phase of the stage
B outburst, hump-like modulations were visible at a period of ${\sim}$
0.048111(354) d. The origin of this period remains unknown.

4. Daily-averaged color-color indices of SU UMa revealed two cycles:
outburst $\rightarrow$ bright quiescence $\rightarrow$ outburst, and
outburst $\rightarrow$ faint quiescence (blue) $\rightarrow$ faint
quiescence (red) $\rightarrow$ outburst. This suggests that the
thermal limit cycle of the accretion disk for SU UMa is more
complicated.

5. The color index in $g'-R_{\rm c}$ becomes redder ${\sim}$ 3 days
prior to an outburst. One possibility is that the inner portion
of the accretion disk is filled by gas, which obscures the central
white dwarf. We cannot rule out the possibility that the stagnation
effect works at the onset of the outburst. The physical mechanism on
this reddening should be elucidated in future observations.

\vskip 5mm
We would like to thank an anonymous referee for helpful comments on
the manuscript of this paper. 
We acknowledge with thanks the variable star observations from the
AAVSO International Database contributed by observers worldwide and
used in this research. A.I. and H.I. are supported by Grant-In-Aid for
Scientific Research (A) 23244038 from Japan Society for the Promotion
of Science (JSPS). This work is partly supported by Optical and
Near-infrared Astronomy Inter-University Cooperation Program,
supported by the MEXT of Japan.

%\begin{figure}
%
%\end{figure}

\end{document}